\documentclass[12pt]{elsart}
\usepackage{graphicx}
\usepackage[dvips]{epsfig}
\usepackage{pstcol,pst-text}
\usepackage{latexsym}
\usepackage{amsfonts}
\usepackage{amsmath}
\def\apa{{\tt APACIC++}}
\def\ame{{\tt AMEGIC++}}
\def\she{{\tt SHERPA}}
\begin{document}
\begin{frontmatter}
\title{Event Generator for Particle Production in High-Energy
       Collisions\thanksref{talk1}}
\thanks[talk1]{Talk given by G.~Soff at the 25th Course of the 
              {\it``International School on Nuclear Physics''} (Erice,
              September 2003)}
\author{A.~Sch{\"a}licke}, %\thanksref{mail1}
\author{T.~Gleisberg},
\author{S.~H{\"o}che},
\author{S.~Schumann},
\author{J.~Winter}, 
\author{F.~Krauss}, and
\author{G.~Soff}, 
\address{Institut f\"ur Theoretische Physik, TU Dresden, 
         01062 Dresden, Germany}
%\thanks[mail1]{E-mail: dreas@theory.phy.tu-dresden.de}
%
\begin{abstract}
Event generators are an indispensable tool for the preparation and
analysis of particle-physics experiments. In this contribution,
physics principles underlying the construction of such computer
programs are discussed. Results, within and beyond the Standard Model
of particle physics, obtained with a new event generator \cite{sherpa}
are presented. This generator is capable to describe signal processes
for exotic physics and their backgrounds at electron--positron and
proton--(anti)proton colliders.
\end{abstract}
\end{frontmatter}
\section{Introduction}
In current collider experiments the hard interaction of two incoming
beams results in the production of up to thousands of outgoing
particles. So far, no evidence has been found that contradicts the
belief that this process is described by the Standard Model (SM) for
strong and electroweak phenomena.
Unfortunately, a full quantum-mechanical treatment is out of reach.
There are two reasons for this apparent shortcoming: first of all, the
sheer number of particles involved gives rise to a tremendous number
of interfering contributions that grows factorially with the number of
particles. Furthermore, perturbation theory is not able to account for
the transition of partons to hadrons. This failure of perturbation
theory necessitates other methods. Simulation tools called Monte Carlo
programs or event generators have proved to be successful for a
detailed description of multiparticle production. In this contribution
we want to briefly illuminate the principles underlying such programs,
and we would like to present some results of our new event generator
\she\ \cite{sherpa}.
\section{Physics in Event Generators\label{basics}}
\begin{figure}[t!]
  \vspace{-1mm}
  \begin{center}
    \includegraphics[width=122mm]{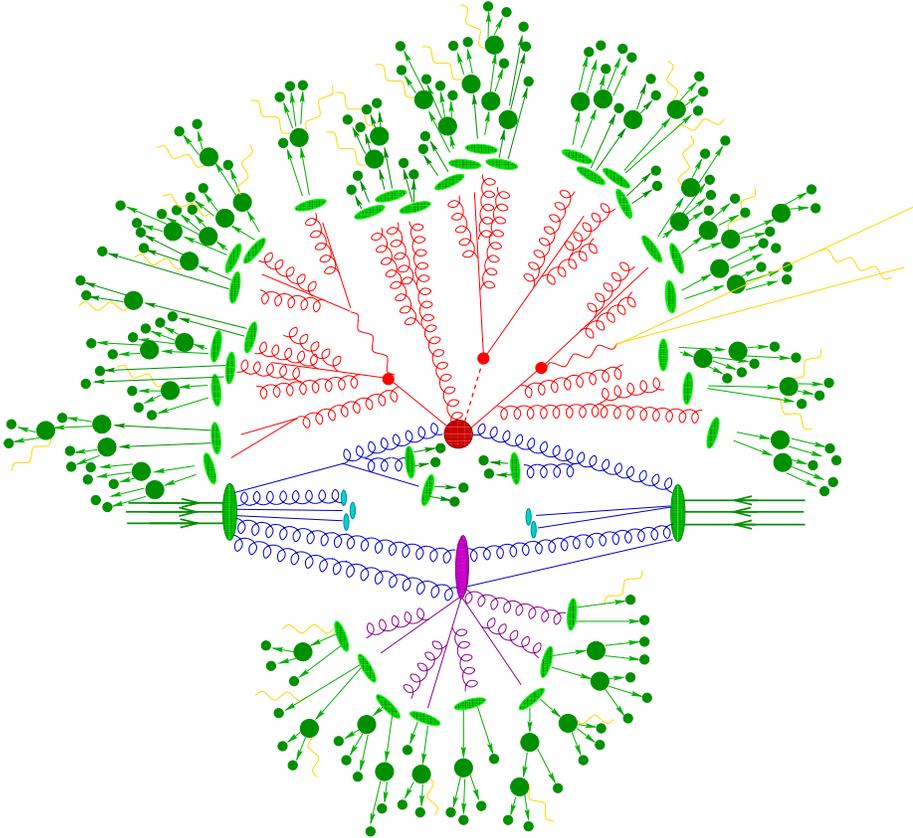}
  \end{center}
  \vspace{-1mm}
  \caption{\label{ppscetch}
           Sketch of a proton--proton collision from an
           event-generator point of view.}
  \vspace{0mm}
\end{figure}
For a brief discussion of the physics concepts underlying event
generators, Fig.~\ref{ppscetch} will be employed, in which a
proton--(anti)proton collision is sketched. The simulation of such an
event has to be decomposed into different independent stages. To
control this independence, these stages are ordered according to the
scales related to them.
\\
At the highest scale the signal process takes place leading eventually
to the production of heavy and/or exotic states. In the figure, this
is represented by the thick blob in the middle, in which two gluons
fuse to produce a four-particle final state. For such signals one
would like to use the full matrix element. Within our event generator
the module \ame\ \cite{ame} takes care of the hard processes. It
automatically creates all Feynman diagrams, stores them as helicity
amplitudes into library files and uses these libraries when the
integral over the phase space of final-state particles, i.e.\ the
cross section, is evaluated through Monte Carlo methods.
\\
As the event evolves down to lower energy scales, multiple emission of
secondary particles must be described. The bulk of radiation activity
is in the soft and/or collinear phase-space region, where the
interference of subsequent emissions is of subleading logarithmic
order only. The logarithms are in ratios of the actual parton-emission
scale compared to a suitably defined harder scale, so that, for a
description of the radiation pattern, these interferences can be
safely neglected. Hence multiple parton emission factorizes into
chains of individual parton branchings. Consequently, one obtains a
simulation in terms of a parton shower. Radiation off the initial
state is more involved, but must result in a boost only, which, due to
recoil effects, may also produce some net transverse momentum of the
final state. In our event generator, the parton showers are performed
by the module \apa\ \cite{apa}, which resums all leading logarithms
and uses the invariant masses of the decaying partons to order
subsequent emissions. An ordering in successive opening angles to
properly account for quantum coherence effects is supplemented. A
novel feature of our code is that, for the first time, an algorithm
has been developed combining matrix elements and the parton shower by
smoothly filling the parton-emission phase space and by respecting the
logarithmic accuracy \cite{CKKW}.
\\
Some particles produced in the hard interaction might be unstable. If
their mass is large enough, their decay might involve particles
emitted with large energies or under large opening angles. Such
decays, again, are treated by using matrix elements and their
subsequent parton showers. At high energies the possibly perturbative
interaction of the beam remnants, sometimes also coined multiple
parton interaction, comes into play. These interactions may develop
their own parton showers. However, this phenomenon has the potential
to severely obscure the interpretation of experimental data and,
therefore, has to be simulated. These two aspects are currently being
implemented.
\\
The gauge structure of QCD leads to a breakdown of perturbation theory
at scales of the order of $\Lambda_{\rm QCD}$. So, the parton shower
has to stop at such scales and its parton ensemble has to be
translated into hadrons, i.e.\ white objects. Since confinement has
not been completely understood yet, hadronization effects have to be
modelled phenomenologically introducing parameters that have to be
fitted to data. Successful models, such as the string and the cluster
model, rely on the fact that before the non-perturbative transition
the parton shower has a definite colour structure%
\footnote{The parton shower is inherently in a large-$N_{\rm C}$
          approximation, where $N_{\rm C}$ is the number of colours.
          Employing a forced splitting of gluons into $q\bar q$\/
          pairs, the partons form a set of colour singlets, each of
          which contains a quark and an antiquark. In cluster models
          usually formulated in terms of formation and decay of
          clusters these singlets are interpreted as primordial hadron
          resonances with indefinite mass decaying into primary,
          specified hadrons.}.
Our code incorporates a newly developed cluster-hadronization model
and an interface to the well-known Lund string fragmentation, allowing
for systematic studies of model uncertainties.
\section{Results\label{results}}
\begin{figure}[h!]
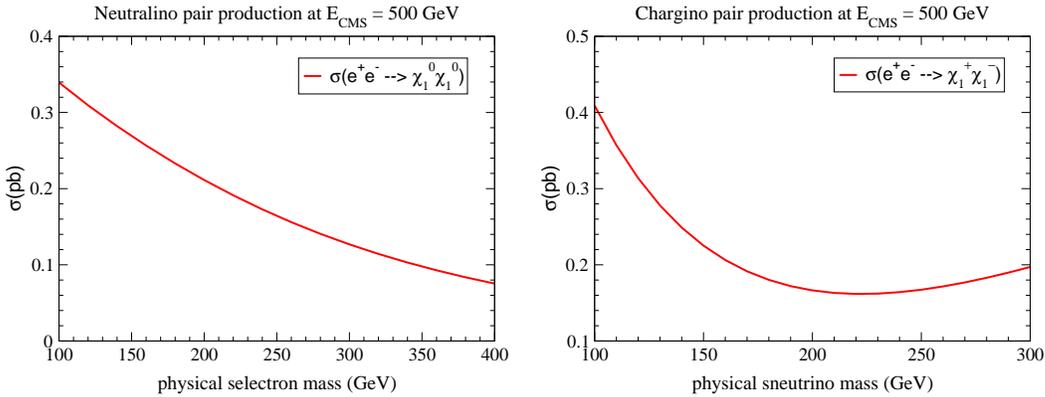

  \vspace{1mm}
  \begin{center}
    \begin{tabular}{cc}
      \hspace*{-2mm}
      \includegraphics[width=67mm]{Chi0SPS1a.eps} &
      \includegraphics[width=67mm]{Chi1SPS1a.eps}
    \end{tabular}
  \end{center}
  \vspace{-1mm}
  \caption{\label{chi0prod}
           $E_{\rm{cm}}=500$ GeV cross sections for
           $e^+e^-\!\to{\chi}_1^0{\chi}_1^0$ (left) and for
           $e^+e^-\!\to\chi_1^+\chi_1^-$ (right) in dependence on the
           selectron and the sneutrino mass, respectively.}
  \vspace{-1mm}
\end{figure}

\begin{figure}[h!]
  \vspace{0mm}
  \begin{center}
    \includegraphics[width=11.7cm]{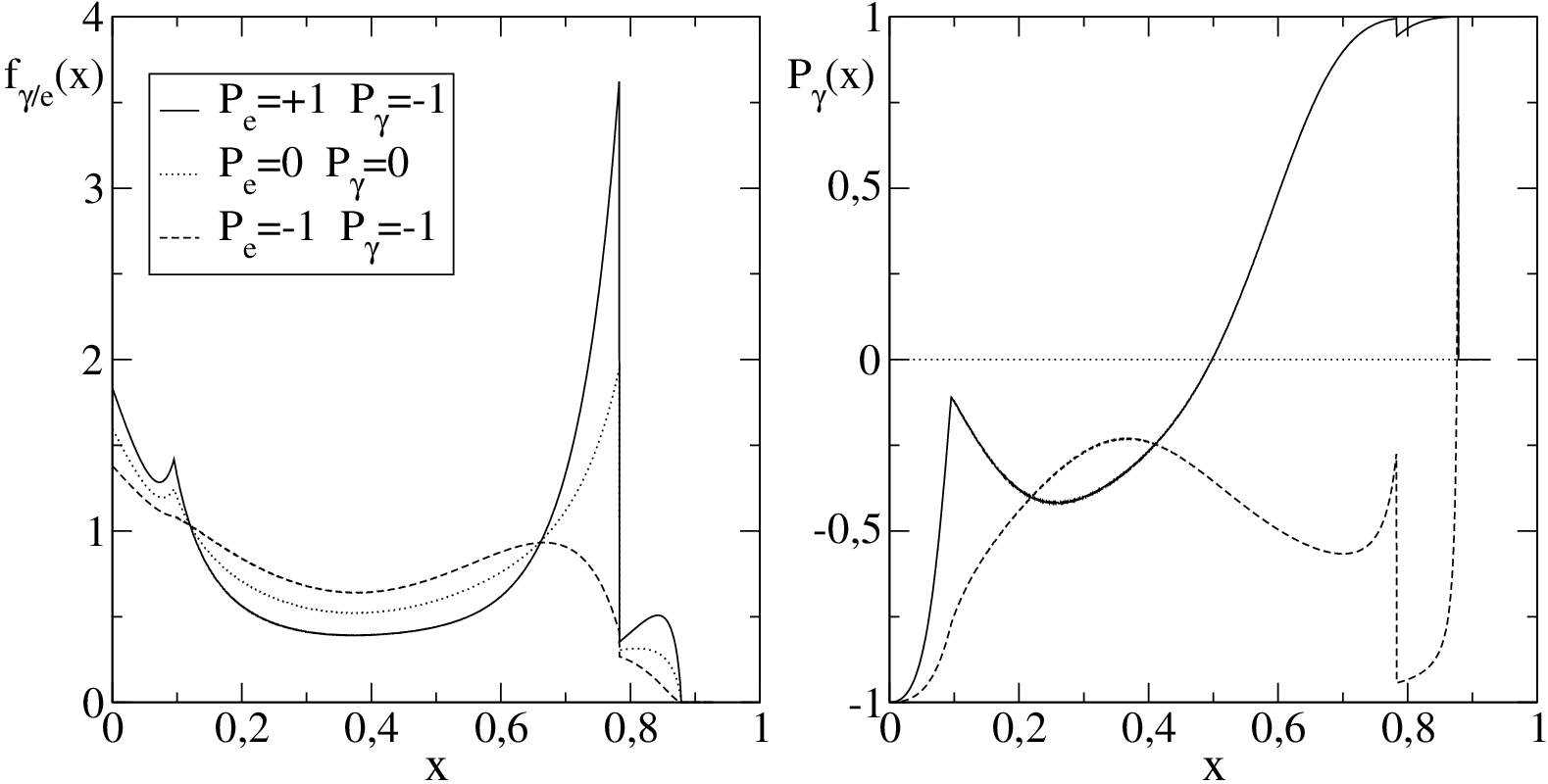}
  \end{center}
  \vspace{-3mm}
  \caption{\label{lbsspec}
      The photon luminosity spectrum and the degree of circular
      polarization as a function of the photon energy-fraction 
      ${\rm x}=E_{\gamma'}/E_e$ for different laser and beam
      helicities in accordance to the parametrization by {\tt CompAZ}
      \cite{compaz}, assuming an electron energy $E_e$ of 250 GeV and
      a laser energy $\omega_\gamma$ of $1.17\cdot 10^{-9}$ GeV.}
  \vspace{3.7mm}
  \begin{center}
    \includegraphics[width=7.7cm]{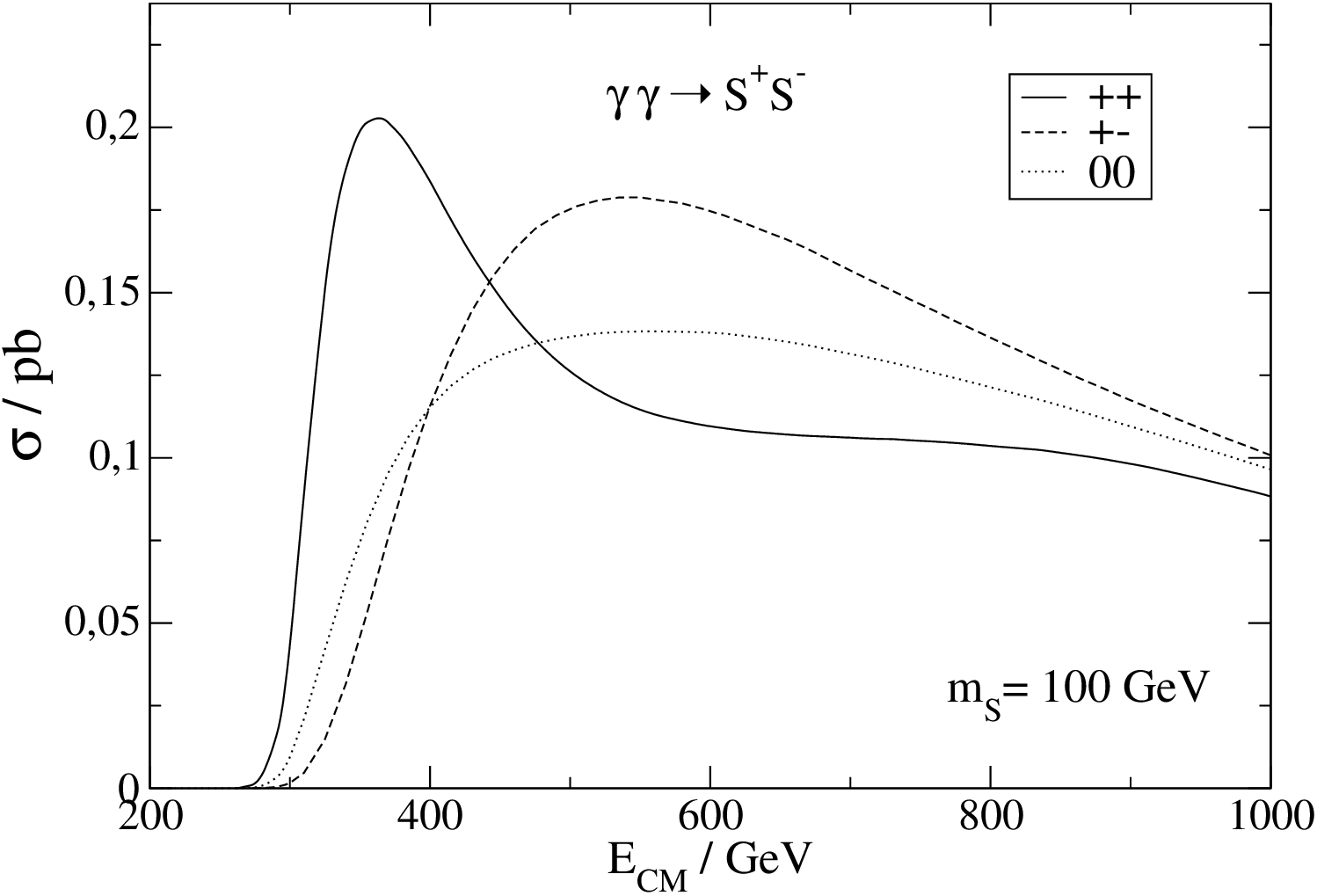}
  \end{center}
  \vspace{-2mm}
  \caption{\label{lbssmuon}
           The total cross section of $\gamma\gamma\to\tilde\mu^+
           \tilde\mu^-$ for $L=0$ ($++$), $L=2$ ($+-$) and completely
           unpolarized beams (00) using the {\tt CompAZ} spectrum,
           note ${\rm S}=\tilde\mu$.}
  \vspace{-1mm}
\end{figure}

\begin{figure}[t]
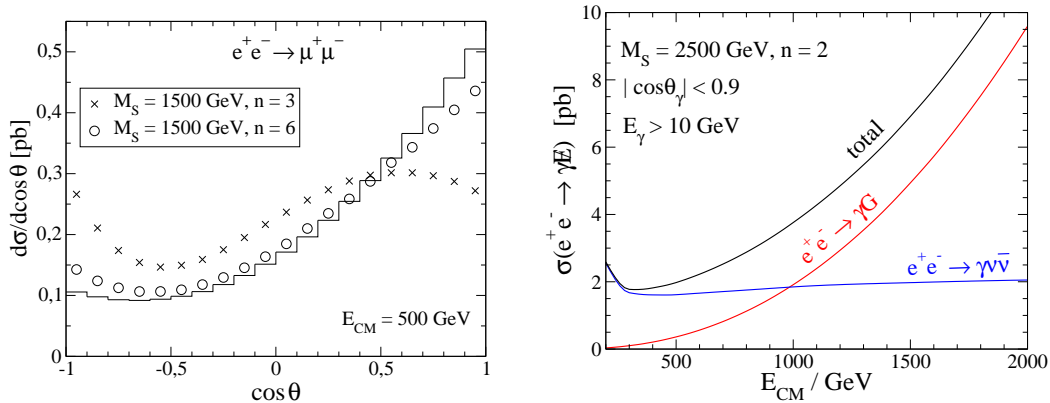

  \vspace{-1mm}
  \begin{center}
    \begin{tabular}{cc}
      \hspace*{-2mm}
      \includegraphics[width=67mm]{diageemumuADD3.eps} &
      \includegraphics[width=67mm]{ee_gG.eps}
    \end{tabular}
  \end{center}
  \vspace{-2.4mm}
  \caption{\label{addmumu_addgammaG}
           a) Differential cross section for $e^+e^-\!\to\mu^+\mu^-$.
           The solid-line histogram shows the SM background. Crosses
           and circles depict virtual graviton exchange in two ADD
           scenarios.
           b) Total cross section for $e^+e^-\!\to\gamma G$ as a
           function of $E_{\rm{cm}}$.}
  \vspace{-1.4mm}
\end{figure}

{\it 3.1\quad Results obtained with \ame}: {\it the parton level}\\[3mm]
Here, we will highlight the abilities of \ame\ in calculating cross
sections for processes within and beyond the SM. There is a plethora
of reasons to search for new physics, e.g.:
\\
The SM has two intrinsic problems, the hierarchy and the naturalness
problem, which are addressed to the question why the scale of
electroweak symmetry breaking and the Planck scale, where strong
gravitational effects are expected, are separated by a wide desert.
To understand this hierarchy self-interactions of the (scalar) Higgs
boson are considered. Using a cut-off procedure for the ultraviolet
divergencies with the cut-off given by the highest scale in the SM,
the Planck mass, one immediately realizes that the counter term for
the Higgs mass has to be quadratic in the Planck mass. In reality,
however, unitarity of $WW$ scattering forces the Higgs mass to be
lower than roughly $1$ TeV. The difference of roughly seventeen orders
of magnitude requiring an extreme fine-tuning (naturalness problem)
lets the SM seem somewhat inconsistent.
\\
On top of this intrinsic problems, gravitation is not incorporated by
the SM. Assuming the existence of a ``theory of everything'', one may
conclude that the SM is incomplete and constitutes only an ``effective
theory''.
\\
However, the hope to successfully search for new physics depends on
our capabilities to understand and subtract the ``old'' physics,
i.e.\ the SM. Thus it is crucial for any event generator to provide an
optimal simulation of SM effects.
\\[2mm]
{\it 3.1.1\quad Multi-particle production in the SM: six-fermion final
states}\\[1mm]
Despite that six-fermion final states constitute significant
backgrounds to signatures for new physics, they provide unique
opportunities to study some of the most interesting aspects of the SM
in great detail. The most important channels are the production and
decay of pairs of top quarks and -- if existent -- of one or more
Higgs bosons, the latter allowing to test the structure of the Higgs
potential. If no evidence for a Higgs boson was found at the LHC, the
study of quartic gauge-boson couplings is mandatory in order to
understand alternative scenarios of electroweak symmetry breaking. To
test these complex calculations involving up to ${\mathcal O}(10^4)$
Feynman diagrams, tuned comparisons of different generators are
required.
\begin{table}[t!]
\begin{center}  
\begin{tabular}{|l||c|c|c|}
\hline
$\sigma$[fb]&{\tt LUSIFER}&{\tt HELAC/PHEGAS}&{\tt AMEGIC++}\\\hline
$\nu_ee^+e^-\bar\nu_eb\bar b$&5.853(7)&5.866(9)&5.879(8)\\[-1mm]
$\nu_ee^+\mu^-\bar\nu_{\mu}b\bar b$&5.819(5)&5.822(7)&5.827(4)\\[-1mm]
$\nu_{\mu}\mu^+\mu^-\bar\nu_{\mu}b\bar b$&5.809(5)&5.809(5)&5.809(5)\\[-1mm]
$\nu_{\mu}\mu^+\tau^-\bar\nu_{\tau}b\bar b$&5.800(4)&5.798(4)&5.800(3)\\[-1mm]
$\nu_{\mu}\mu^+d\bar ub\bar b$&17.171(24)&17.204(18)&17.209(9)\\\hline
\end{tabular}
\vspace{-2mm}
\caption{\label{sixferm}%
         Comparison of tree-level predictions corresponding to
         six-fermion top-quark pair-production channels at
         $E_{\rm{cm}}=0.5$ TeV for massless fermions; see
         \cite{Dittmaier:2003sc}.}
\end{center}
\vspace{-1mm}
\end{table}
Table \ref{sixferm} presents predictions of {\tt LUSIFER}
\cite{Dittmaier:2002ap}, {\tt HELAC/PHEGAS} \cite{Kanaki:2000ey} and
\ame\ \cite{ame} for some six-fermion final states corresponding to
the production and decay of a top-quark pair. More extensive
comparisons of results obtained by {\tt HELAC/PHEGAS} and \ame\ also
taking into account finite fermion masses can be found in
\cite{Gleisberg:2003}.
\\[1mm]
{\it 3.1.2\quad Cross sections for Supersymmetry}\\[1mm]
The ability of our code for searches of new physics is exemplified by
some processes within the framework of the MSSM -- the Minimal
Supersymmetric extension of the Standard Model, see e.g.\ \cite{susy}.
Supersymmetry is one of the most promising candidates for physics
beyond the SM, since it cures the hierarchy problem in a natural
fashion: it predicts at least one superpartner for each particle, a
bosonic (fermionic) partner for a fermion (boson), with identical
quantum numbers and mass. The only difference is in the spins, which
differ by one half. In loops, the partners come with opposite sign,
thus the quadratic divergencies mentioned before are cancelled.
However, no superpartner has been found yet. Hence, Supersymmetry must
be broken. The demand to solve the hierarchy problem implies a soft
breaking, i.e.\ the form of breaking terms is limited. First, one
usually considers a minimal Supersymmetry, i.e.\ one superpartner per
particle of the SM, and this MSSM is broken by adding all possible
breaking terms.
\\
The complete MSSM has been implemented in \ame, and an example is
presented in Fig.~\ref{chi0prod}: gaugino-pair production at an
$e^+e^-$ collider shows a strong dependence on the masses of the
sleptons propagating in the t-channel. This is true for the lightest
charginos ($\chi_1^\pm$) and neutralinos ($\chi_1^0$). Starting with
the mSUGRA point%
\footnote{For current studies of the phenomenology of the MSSM,
          specific benchmark points have been proposed
          \cite{spspoints}.}
SPS1a \cite{spspoints}, the appropriate slepton masses have been
varied. For the charginos, the rise for large sneutrino mass exhibits
the vanishing destructive interference of s- and t-channel diagrams.
\\
Besides $e^+e^-$ physics, a next linear collider might provide an
unique opportunity to study $\gamma\gamma$ and $\gamma e$ interactions
at high luminosities. Both can be simulated within our event
generator. High-energetic electrons are transformed through Compton
backscattering of laser light into high-energy photons. This energy
spectrum is shown in the left part of Fig.~\ref{lbsspec}. Using the
polarization dependence of the backscattered photons, valuable
information on particles produced in collisions of polarized beams can
be obtained. The right part of Fig.~\ref{lbsspec} depicts the photon
polarization in dependence on its energy-fraction w.r.t. the incoming
electron. Figure~\ref{lbssmuon} illuminates smuon pair production
exhibiting a significant dependence on the polarization states of the
$\gamma$ beams.
\\[1mm]
{\it 3.1.3\quad Cross sections for Large Extra Dimensions}\\[1mm]
An alternative physics scenario has been recently put forward by
Arkani-Hamed, Dimopoulos, and Dvali (ADD model) \cite{ADD}. They
propose to solve the hierarchy problem of the Standard Model by
shifting the Planck scale down to scales of the order of TeV. This is
achieved by introducing $n\ge2$ new compact spatial dimensions that
are large compared to the scale of electroweak symmetry breaking.
These dimensions can in principle be as large as fractions of
millimeters, since only down to these length scales gravity has been
probed experimentally. To ensure that the SM remains valid within the
regions tested so far, SM particles remain constrained to the common
four-dimensional manifold. Only gravitons propagate in all dimension.
This changes the laws of gravity and leads to a number of striking
signals for the next generation of collider experiments.
To study the rich phenomenology of this ADD model, a correct treatment
of spin-2 particles in the helicity formalism has been developed and
implemented \cite{Gleisberg:2003ue}. This allows to describe real as
well as virtual graviton production, both of which exemplified in
Fig.~\ref{addmumu_addgammaG}.
\\[4mm]
{\it 3.2\quad Results obtained at the hadron level}\\[3mm]
They are presented for $e^+e^-$ collisions at the $Z^0$ pole, $E_{\rm
cm}=91.2$ GeV. We will focus on the overall performance of our event
generator in its ability to describe multi-hadron production, and on
some aspects of our cluster model.
\\[2mm]
{\it 3.2.1\quad Overall performance}\\[1mm]
First, inclusive observables are considered.
\begin{figure}[t!]
  \vspace{-7mm}
  \begin{center}\hspace*{-4mm}
    \includegraphics[width=72mm]{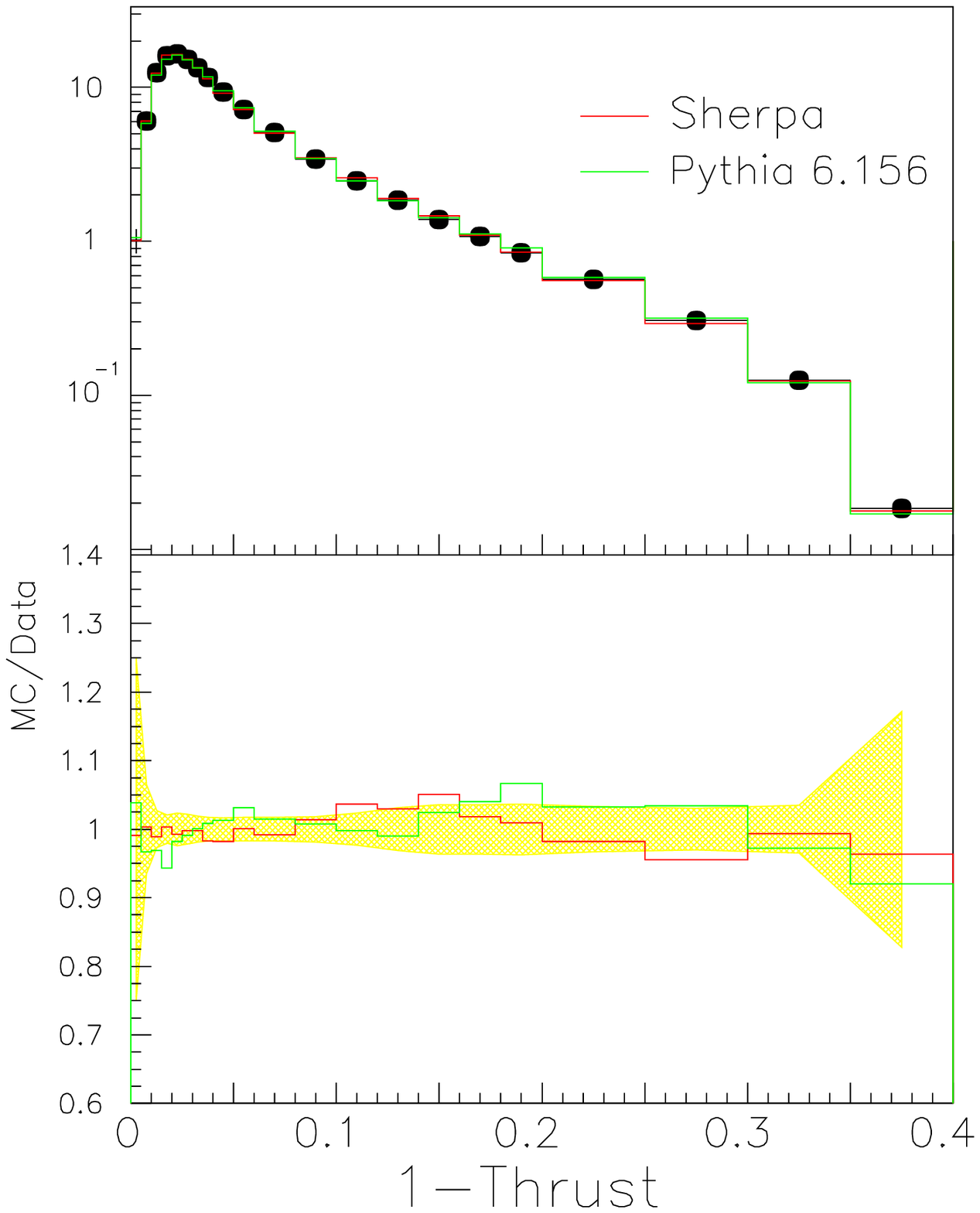}\hspace*{-3mm}
    \includegraphics[width=72mm]{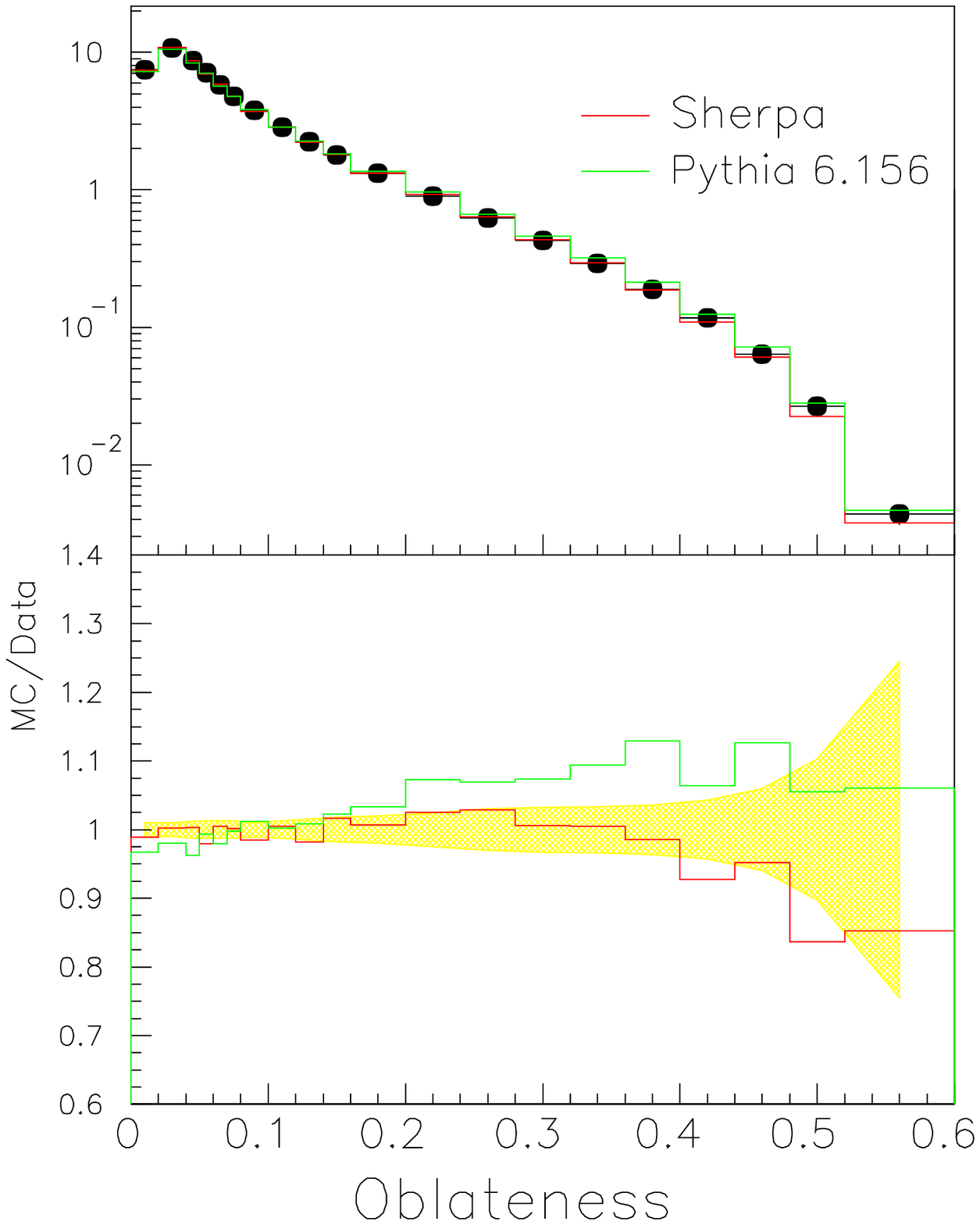}
  \end{center}
  \vspace{-8mm}
  \caption{\label{thrust}
           Event-shape variables for $e^+e^-\!\to\rm{hadrons}$ at the
           $Z^0$ pole.}
  \vspace{-1mm}
  \begin{center}\hspace*{-4mm}
    \includegraphics[width=72mm]{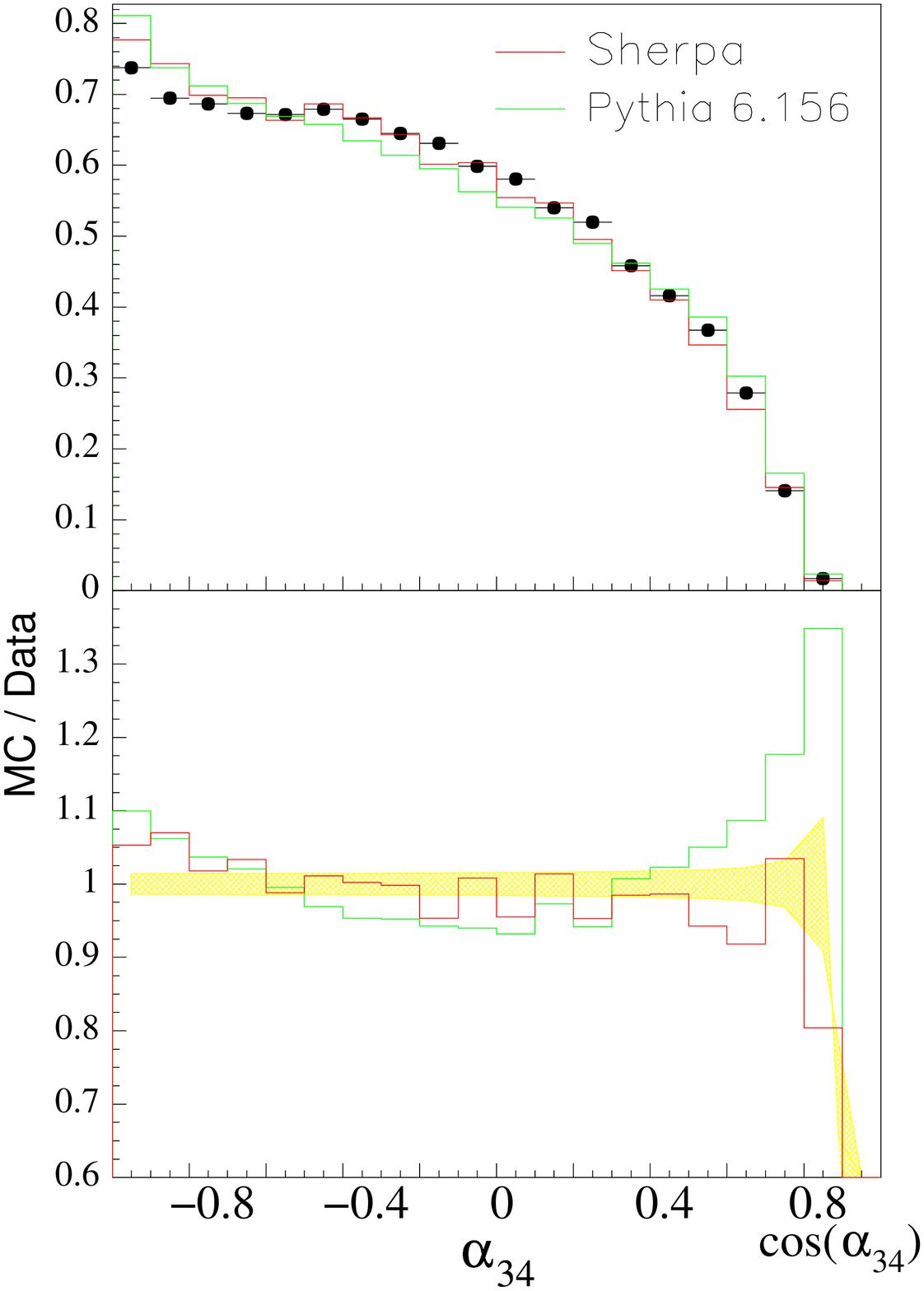}\hspace*{-3mm}
    \includegraphics[width=72mm]{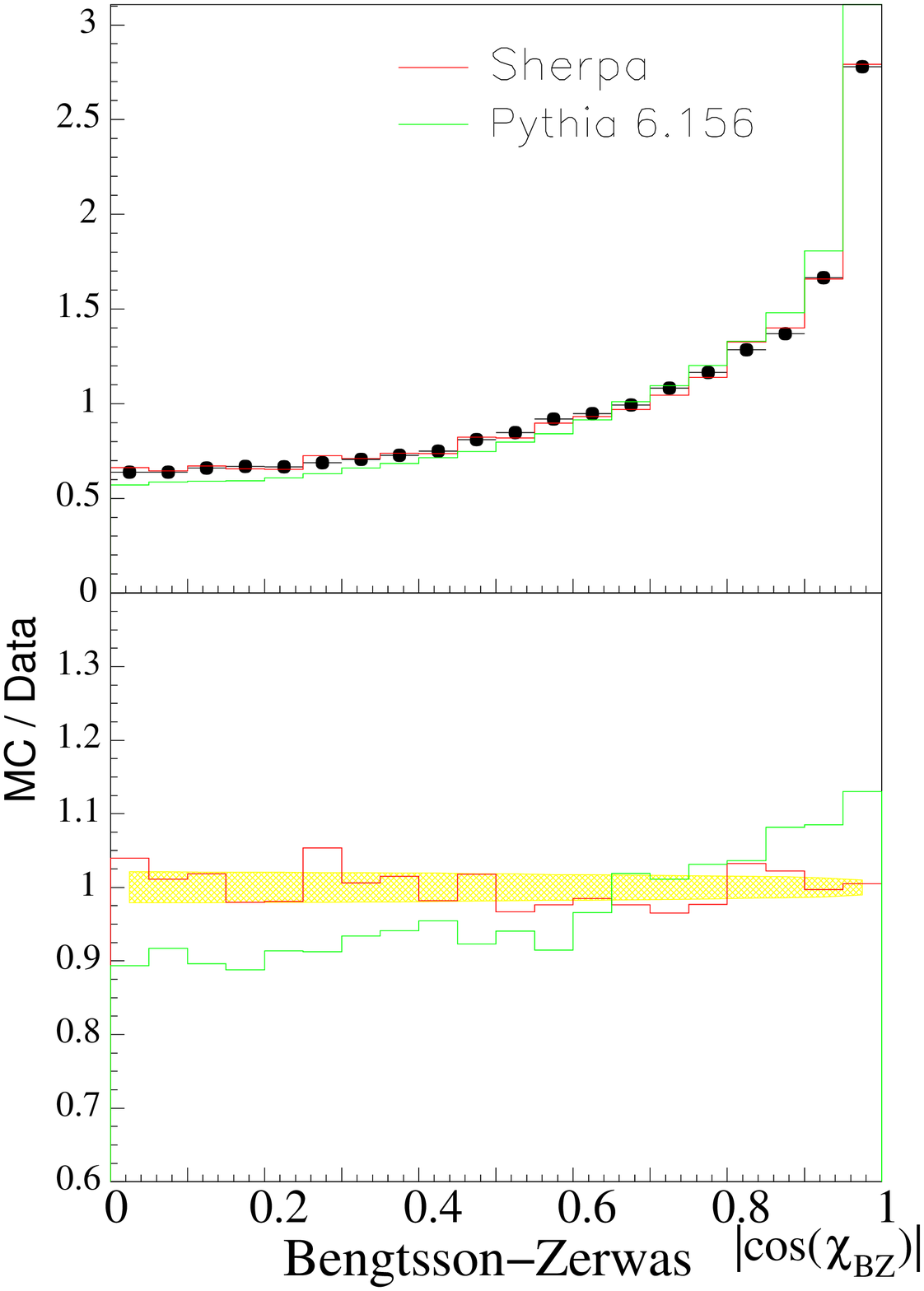}
  \end{center}
  \vspace{-7mm}
  \caption{\label{alphaBZ}\label{alpha34}
           The $\alpha_{34}$ and the Bengtsson--Zerwas angle
           \cite{Bengtsson:1988qg} for QCD four-jet events.}
  \vspace{-2mm}
\end{figure}
This is usually done using event-shape variables, such as thrust,
sphericity, oblateness etc., or by looking into the correlations of
hadronic matter -- jets -- deposed in the detector, such as relative
angles of jets.
In Fig.~\ref{thrust} oblateness and $1-$thrust are depicted. The
results of our new generator \she\ \cite{sherpa} are confronted with
the results of the well-known {\tt PYTHIA} Monte-Carlo \cite{Pythia}
and with experimental data obtained by ALEPH \cite{Barate:1996fi}.
Both predictions agree well with the data. Moreover, using DELPHI data
\cite{hoeth}, Fig.~\ref{alphaBZ} shows the comparison for two more
exclusive observables, the $\alpha_{34}$ and the Bengtsson--Zerwas
relative angle in four-jet events. There, the impact of our new
approach to merge matrix elements and parton showers proves its
predictive power. In contrast, {\tt PYTHIA} adds radiation to the
initial two-jet configuration mainly through the parton shower.
Therefore, it includes less quantum interference effects. This is the
source of the deviation.
\\[2mm]
{\it 3.2.2\quad Cluster-hadronization model}\\[1mm]
Inspired by the well established Webber model \cite{Webber:1983if} for
primary hadronization of a coloured partonic system, we have developed
a new model presented in \cite{Winter:2003tt} and extended by the
following features: soft colour reconnection is modelled and
incorporated in the formation and fragmentation of clusters, yielding
additional cluster-decay configurations. A new mesonic cluster type,
the two-diquark cluster, and its decay channels are added. The
explicit consideration of the spin of diquark cluster-constituents is
rendered possible. The most significant feature of our treatment is
the flavour-dependent separation of the cluster and hadron regimes in
terms of the cluster's mass. As a consequence, the different
cluster-transition modes are obtained automatically.
\begin{figure}[t!]
  \vspace{-2mm}
    \begin{center}\includegraphics[height=57mm]{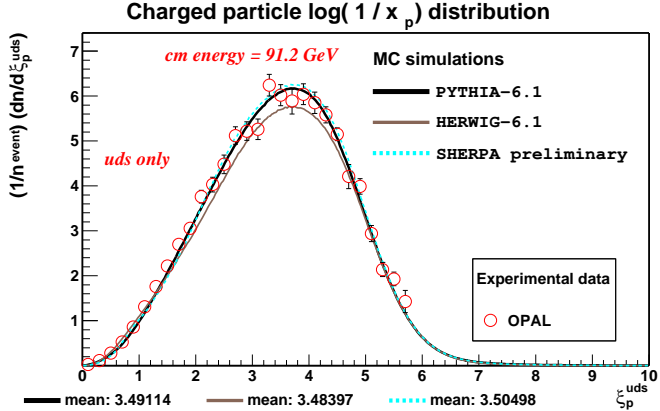}\end{center}
  \vspace{-3mm}
  \caption{$\xi^{uds}_p=\ln(1/x^{uds}_p)$ distribution of charged
           particles for $E_{\rm{cm}}=91.2$ GeV in $e^+e^-$
           annihilation into light quarks only. The {\tt preliminary\
           SHERPA} prediction is presented together with OPAL data
           \cite{Ackerstaff:1998hz}, and with results from default
           {\tt PYTHIA-6.1($uds$)} and default {\tt HERWIG-6.1($uds$)}
           \cite{Corcella:1999qn}.}
  \label{xip}
  \vspace{0mm}
\end{figure}
\\
A {\tt preliminary SHERPA} version including our cluster-hadronization
model is capable to describe $e^+e^-\!\to\gamma^{\ast}/Z^0\!\to d\bar
d, u\bar u, s\bar s\to\rm{hadrons}$. The agreement with default {\tt
PYTHIA-6.1($uds$)} \cite{Pythia} and experimental light-quark data is
satisfactory. To illustrate this, the distribution of the negative
logarithm of the scaled-momentum, $\xi^{uds}_p=-\log x^{uds}_p$, is
shown in Fig.~\ref{xip}.
\section{Conclusion\label{concl}}
Event generators will continue to be an indispensable tool for future
experimental analyses. A new multi-purpose generator for high-energy
collisions, \she\ \cite{sherpa}, is currently being developed. It
already proved to successfully describe $e^+e^-$ annihilation. The
embedded automated matrix-element calculator \ame\ \cite{ame} is not
only capable to evaluate all Standard Model processes, but also
physics beyond the Standard Model (e.g.\ MSSM and ADD).
\\
The code is written in a transparent structure using the
object-oriented programming language {\tt C++}. Consequently, the
implementation of extensions is straightforward; the code is easy to
maintain and adjust to the needs of users.
\\
\\
We acknowledge that this work has been supported by BMBF and GSI.

\end{document}